\documentclass[titlepage,twoside,12pt]{article}
\usepackage{amssymb}
\usepackage{amsfonts}
\begin{document}

{\LARGE \bf A General Scheme of Entanglement} \\ \\


{\bf Elem\'{e}r E ~Rosinger} \\ \\
{\small \it Department of Mathematics \\ and Applied Mathematics} \\
{\small \it University of Pretoria} \\
{\small \it Pretoria} \\
{\small \it 0002 South Africa} \\
{\small \it eerosinger@hotmail.com} \\ \\

{\bf Abstract} \\

Entanglement is a well known fundamental resource in quantum information. Here the following
question is addressed : which are the deeper roots of entanglement that may help in its better
understanding and use ? The answer is that one can reproduce the phenomenon of entanglement in
a far more general and simple way, a way that goes much beyond the usual one which is limited
to the framework of tensor products of vector spaces. In this general approach to entanglement
presented here - and much unlike in the particular setup of tensor products of vector spaces - the
spaces involved can be rather arbitrary sets, just as in the case of Cartesian products. In
particular, they need not even have any algebraic structure. Thus they do not have to be vector
spaces, groups or even semigroups. \\ \\ \\

{\bf 0. Preliminaries} \\

Entanglement, as an essential physical phenomenon in Quantum Mechanics, appeared for the first
time in the celebrated 1935 EPR paper of Einstein, Podolsky and Rosen, [2]. The term {\it
entanglement} itself, however, was not used in that paper. Instead, it was introduced, in
German, in the subsequent papers of Schr\"{o}dinger, [4,5], in which he commented on the state of
Quantum Mechanics at the time, and among others, brought to attention the problematic
situation which ever since would be called {\it Schr\"{o}dinger's cat}. \\
As for the special place of the EPR paper in Quantum Mechanics, suffice it to mention that
till the 1980s, it has already been cited more than one million times, [1], and even today,
more than seven decades after its publication, it is still the paper most often downloaded
from the web site of the American Physical Society, [6]. \\

Independently, and prior to that, in Multilinear Algebra, the concept of {\it tensor product}
was introduced by mathematicians in view of its {\it universal} property of establishing a
natural connection between multilinear and linear mappings, see Appendix. \\

It took some time, both for physicists and mathematicians, to become aware of the fact that a
natural mathematical formulation of quantum entanglement can be obtained with tensor products
of Hilbert spaces. And this issue has lately become even more fundamental with the emergence
and massive development of Quantum Information, and in particular, Quantum Computation, where
entanglement proves to be one of the most important quantum resources that can make quantum
computers significantly more powerful than the usual digital ones. \\

Indeed, let us recall that within Classical Mechanics, given two systems $S$ and $S\,'$ whit
the respective state spaces $X$ and $X\,'$, their composition will have the state space given
by the {\it Cartesian product} $X \times X\,'$. \\
On the other hand, in Quantum Mechanics, the composition of two systems $Q$ and $Q\,'$, with
the respective state spaces $H$ and $H\,'$, is given by the {\it tensor product} $H \bigotimes
H\,'$ of the Hilbert spaces $H$ and $H\,'$. \\

And to have a first and simple appreciation of the difference, let us recall that given two
vector spaces $E$ and $F$, with the respective finite dimensions $m$ and $n$, the dimension of
their Cartesian product $E \times F$ is $m + n$, while that of their tensor product $E
\bigotimes F$ is $m n$. \\
In Quantum Computation this difference has the following dramatic convenient effect. The basic
quantum information, a qubit, is an element of the two dimensional Hilbert space $H =
\mathbb{C}^2$. Consequently, the state space of an $n$-qubit quantum system, a typical
register in a quantum computer, is described by the $n$-factor tensor product $\mathbb{C}^2
\bigotimes \ldots \bigotimes \mathbb{C}^2$ which has the dimension $2^n$. Therefore, one can
obtain an {\it exponentially} fast increase in the capacity of a quantum register, simply by a
linear increase in the length of that quantum register. Needless to say, there is nothing
similar possible with usual digital computers, where the composition giving registers happens
in the realms of Classical Mechanics, hence through Cartesian products. \\

In this way, the essence of entanglement is that it represents the {\it essential difference} between
the Cartesian product, and on the other hand, the tensor product, making the latter far larger.
Consequently, a study of the more general roots of entanglement can further clarify to what extent
the quantum way of composing systems, that is, by tensor product, can indeed lead to far more rich
systems than the classical way of system composition which is by Cartesian product. \\
And it is not a priori impossible that a formulation of entanglement in a more general framework, such as
for instance pursued here, may even lead to "new physics" related to quantum phenomena. \\

The aim of this paper, in view of the above, is as follows. So far, entanglement was modelled
mathematically only by tensor products of vector spaces. In view of the significant importance
of entanglement, one can ask the question :

\begin{itemize}

\item Can entanglement be modelled in other more {\it general} ways, than by tensor products
of vector spaces ?

\end{itemize}

Here we give an affirmative answer to this question, by presenting general, and yet simple
ways of entanglement, which contain as a particular case tensor products. In fact, in that
general approach to entanglement, and unlike in the particular case of tensor products of
vector spaces, the spaces involved can be rather arbitrary sets, just as in the case of
Cartesian products, thus in particular, they need {\it not} be vector spaces, and not even
groups or semigroups. \\ \\

{\bf 1. Generators and Bases} \\

{\bf Definition 1.1.} \\

Given any set $X$, a mapping $\psi : {\cal P} ( X ) \longrightarrow {\cal P} ( X )$ will be
called a {\it generator}, if and only if \\

(1.1)~~~ $ \forall~~~ A \subseteq X ~:~ A \subseteq \psi ( A ) $ \\

and \\

(1.2)~~~ $ \forall~~~ A \subseteq A\,' \subseteq X ~:~ \psi ( A ) \subseteq \psi ( A\,' ) $ \\

{\bf Examples 1.1.} \\

1) A trivial example of generator is given by $\psi = id_{{\cal P} ( X )}$, that is, $\psi
( A ) = A$, for $A \subseteq X$. \\

2) An example which is important in the sequel is obtained as follows. Given any binary
operation $\alpha : X \times X \longrightarrow X$, we call a subset $A \subseteq X$ to be
$\alpha$-{\it stable}, if and only if \\

(1.3)~~~ $ x, y \in A ~~\Longrightarrow~~ \alpha ( x, y ) \in A $ \\

Obviously, $X$ is $\alpha$-stable, and the intersection of any family of $\alpha$-stable
subsets is $\alpha$-stable. Consequently, for every subset $A \subseteq X$, we can define the
smallest $\alpha$-stable subset which contains it, namely \\

(1.4)~~~ $ [ A ]_\alpha = \bigcap_{A \subseteq B,~ B ~\alpha-stable}~ B $ \\

Therefore, we can associate with $\alpha$ the mapping  $\psi_\alpha : {\cal P} ( X )
\longrightarrow {\cal P} ( X )$ defined by \\

(1.5)~~~ $ \psi_\alpha ( A ) = [ A ]_\alpha,~~~ A \subseteq X $ \\

which is obviously a generator. Furthermore, we have in view of (1.4) \\

(1.6)~~~ $ \forall~~~ A \subseteq X ~:~
       \psi_\alpha ( \psi_\alpha ( A ) ) = \psi_\alpha ( A ) $ \\

since as mentioned, $[ A ]_\alpha$ is $\alpha$-stable, and obviously $[ A ]_\alpha \subseteq
[ A ]_\alpha$. \\

We note that, in general, the relation $\psi ( \psi ( A ) ) = \psi ( A )$, with $A \subseteq
X$, need not hold for an arbitrary generator $\psi$. \\

3) A particular case of 2) above is the following. Let $( S,\ast )$ be a semigroup with the
neutral element $e$. Then $[ \{ e \} ]_\ast = \{ e \}$, while for $a \in S,~ a \neq e$, we
have $[ \{ a \} ]_\ast = \{ a, a \ast a, a \ast a \ast a, \dots \}$. \\

For instance, if $( S, \ast ) = ( \mathbb{N}, + )$, then $[ \{ 1 \} ]_+ = \mathbb{N} \setminus
\{ 0 \} = \mathbb{N}_1$. \\

{\bf Definition 1.2.} \\

Given a generator $\psi : {\cal P} ( X ) \longrightarrow {\cal P} ( X )$, a subset $B
\subseteq X$ is called a $\psi$-{\it basis} for $X$, if and only if \\

(1.7)~~~ $ \psi ( B ) = X $ \\

Let us denote by \\

(1.8)~~~ $ {\cal B}_\psi ( X ) $ \\

the set of all $B \subseteq X$ which are a $\psi$-basis for $X$. In view of (1.1), obviously
$X  \in {\cal B}_\psi ( X )$. \\

{\bf Note 1.1.} \\

1) In view of 3) in Examples 1.1., it follows that neither $\{ 0 \}$, nor $\{ 1 \}$ are
$\psi_+$-bases in $( \mathbb{N}, + )$, while on the other hand, $\{ 0, 1 \}$ is. \\

2) Given a binary operation $\alpha : X \times X \longrightarrow X$, then in view of (1.5),
a subset $B \subseteq X$ is a $\psi_\alpha$-basis for $X$, if and only if \\

(1.9)~~~ $ [ B ]_\alpha = X $ \\

{\bf Definition 1.3.} \\

Given a generator $\psi$ on a set $X$. A binary operation $\alpha$ on $X$ is called {\it compatible} with $\psi$, if and
only if, see (1.5) \\

(1.10)~~~ $ \psi_\alpha ( A ) \subseteq \psi ( A ),~~~ A \subseteq X $ \\

{\bf Remark 1.1.} \\

1) Obviously, $\alpha$ is compatible with $\psi_\alpha$, for every binary operation $\alpha$ on $X$. \\

2) Given any generator $\psi$ on a set $X$, there exist at least two binary operations $\lambda_X$ and $\rho_X$ which are compatible with $\psi$,
namely, given by \\

(1.11)~~~ $  \lambda_X ( x, x\,' ) = x,~~ \rho_X ( x, x\,' ) = x\,',~~~ x, x\,' \in X $ \\

since we obviously have \\

(1.12)~~~ $ \psi_{\lambda_X} ( A ) = \psi_{\rho_X} ( A ) = A \subseteq \psi ( A ),~~~ A \subseteq X $ \\

thus \\

(1.13)~~~ $ \psi_{\lambda_X} = \psi_{\rho_X} = id_{{\cal P} ( X )} $ \\ \\

{\bf 2. Covering Generators} \\

{\bf Definition 2.1.} \\

Given the sets $X$ and $Y$, with the corresponding generators $\psi : {\cal P} ( X )
\longrightarrow {\cal P} ( X )$, $\varphi : {\cal P} ( Y ) \longrightarrow {\cal P} ( Y )$,
and $\chi : {\cal P} ( X \times Y ) \longrightarrow {\cal P} ( X \times Y )$. We say that $\chi$
is {\it covered} by $\psi, \varphi$, if and only if \\

(2.1)~~~ $ \forall~~~ A \subseteq X,~ B \subseteq Y ~:~
                    \chi ( A \times B ) \subseteq \psi ( A ) \times \varphi ( B ) $ \\

{\bf Example 2.1.} \\

Obviously, if $\psi = id_{{\cal P} ( X )},~ \varphi = id_{{\cal P} ( Y )}$ and $\chi =
id_{{\cal P}( X \times Y )}$, then $\chi$ is a covering for $\psi, \varphi$.

\hfill $\Box$ \\

Let now $\alpha : X \times X \longrightarrow X$ and $\beta : Y \times Y \longrightarrow Y$ be
two binary operations on $X$, respectively, $Y$, and, as usual, let us associate with them the
binary operation $\alpha \times \beta : ( X \times Y ) \times ( X \times Y ) \longrightarrow
( X \times Y )$ on $X \times Y$, given by \\

(2.2)~~~ $ ( \alpha \times \beta ) ( ( x, y ), ( u, v ) ) =
             ( \alpha ( x, u ), \beta ( y, v ) ),~~~ x, u \in X,~ y, v \in Y $ \\

Then (1.5) results in \\

{\bf Lemma 2.1.} \\

1) Given $A \subseteq X,~ B \subseteq Y$. If $A$ is $\alpha$-stable and $B$ is $\beta$-stable, then
$A \times B$ is $\alpha \times \beta$-stable. \\

2) $\psi_{\alpha \times \beta}$ is covered by $\psi_\alpha,~ \psi_\beta$, since we have \\

(2.3)~~~ $ [ A \times B ]_{\alpha \times \beta} \subseteq [ A ]_\alpha \times [ B ]_\beta $ \\

{\bf Proof.} \\

1) Let $( x, y ), ( u, v ) \in A \times B$, then $ ( \alpha \times \beta ) ( ( x, y ), ( u, v ) ) =
( \alpha ( x, u ), \beta ( y, v ) ) \in A \times B$. Indeed, $x, u \in A$, thus $\alpha ( x, u ) \in A$,
since $A$ is $\alpha$-stable. Similarly $\beta ( y, v ) \in B$. \\

2) In view of (1.4), $A \subseteq [ A ]_\alpha,~ B \subseteq [ B ]_\beta$. Hence \\

$~~~~~~ A \times B \subseteq [ A ]_\alpha \times [ B ]_\beta $ \\

Now 1) above, yields $[ A ]_\alpha \times [ B ]_\beta$ is $\alpha \times \beta$-stable, hence (1.4) gives \\

$~~~~~~ [ A \times B ]_{\alpha \times \beta} \subseteq [ A ]_\alpha \times [ B ]_\beta $ \\

which in view of (2.1) completes the proof. \\

{\bf Remark 2.1.} \\

Related to (2.3), let $( X, \alpha ) = ( Y, \beta ) = ( S, \ast )$, where $\ast$ is any binary operation on $S$. Then $( X \times Y, \alpha \times
\beta ) = ( S^2, \diamond )$, where \\

(2.4)~~~ $ ( x, y ) \diamond ( u, v ) = ( x \ast u, y \ast v ),~~~ x, u, y, v \in S $ \\

and we note that, in general, we have \\

(2.5)~~~ $ [ A \times B ]_\diamond \subsetneqq [ A ]_\ast \times [ B ]_\ast$ \\

Indeed, let $( X, \alpha ) = ( Y, \beta ) = ( \mathbb{N}, + )$, while $A = B = \{ 1 \}$. Then $[ A ]_+ = [ B ]_+ = \mathbb{N}_1$, thus $[ A ]_+ \times
[ B ]_+ = \mathbb{N}_1^2$. On the other hand, $ ( X \times Y, \alpha \times \beta ) = ( \mathbb{N}_1^2, \oplus )$, where $( x, y ) \oplus ( u, v ) =
( x + u, y + v )$, for $x, u, y, v \in \mathbb{N}$. Furthermore, $A \times B = \{ ( 1, 1 ) \}$, hence $[ A \times B ]_\oplus = \{ ( n, n ) ~|~ n \in
\mathbb{N}_1 \}$. \\ \\

{\bf 3. A First More General Case of Entanglement} \\

Let us present a first generalization of the standard definition of {\it tensor products}, see Appendix, namely, for the case of two structures
$( X, \alpha )$ and $( Y, \beta )$, where $\alpha : X \times X \longrightarrow X,~ \beta : Y \times Y \longrightarrow Y$ are arbitrary binary operations
on arbitrary given sets $X$ and $Y$, respectively. One way to proceed, convenient in the sequel, is as follows. Let us denote by $Z$ the set of all
finite sequences of pairs \\

(3.1)~~~ $ ( x_1, y_1 ), \dots , ( x_n, y_n ) $ \\

where $n \in \mathbb{N}_1$, while $x_i \in X,~ y_i \in Y$, with $1 \leq i \leq n$. We define on $Z$ the binary operation $\gamma$ simply by the
concatenation of the sequences (3.1). It follows that $\gamma$ is associative, therefore, each sequence (3.4) can be written as \\

(3.2)~~~ $ ( x_1, y_1 ), \dots , ( x_n, y_n ) =
                 ( x_1, y_1 ) \gamma ( x_2, y_2 ) \gamma \ldots \gamma ( x_n, y_n ) $ \\

where for $n = 1$, the right hand term is understood to be simply $( x_1, y_1 )$. Obviously, if $X$ or $Y$ have at least two elements, then $\gamma$ is
not commutative. Thus (3.1), (3.2) give \\

(3.3)~~~ $ Z = \left \{ ( x_1, y_1 ) ~\gamma~ ( x_2, y_2 ) ~\gamma~ \ldots ~\gamma~ ( x_n, y_n ) ~~
                             \begin{array}{|l}
                              ~ n \geq 1 \\ \\
                              ~ x_i \in X,~ y_i \in Y,~ 1 \leq i \leq n
                             \end{array} \right \} $ \\ \\

which obviously gives \\

(3.4)~~~ $ X \times Y \subseteq Z $ \\

Now we define on $Z$ an equivalence relation $\approx_{\alpha, \beta}$ as follows. Two sequences in (3.1) are equivalent, if and only if they are
identical, or each can be obtained from the other by a finite number of applications of the following operations : \\

(3.5) permuting the pairs $( x_i, y_i )$ within the sequence \\

(3.6) replacing $( \alpha ( x_1, x\,'_1 ) , y_1 ), ( x_2, y_2 ), \ldots , ( x_n, y_n )$ with \\
      \hspace*{0.8cm} $( x_1, y_1 ), ( x\,'_1, y_1 ), ( x_2, y_2 ), \ldots , ( x_n, y_n )$, or
      vice-versa \\

(3.7) replacing $( x_1, \beta ( y_1, y\,'_1 ) ), ( x_2, y_2 ), \ldots ,( x_n, y_n )$ with \\
       \hspace*{1cm} $( x_1, y_1 ), ( x_1, y\,'_1 ), ( x_2, y_2 ), \ldots , ( x_n, y_n )$, or
       vice-versa \\

Let us note that, in view of the rather general related result in Lemma 3.1. at the end of this section, the binary
relation $\approx_{\alpha, \beta}$ defined above on $Z$ is indeed an equivalence relation. \\

Finally, the {\it tensor product} of $( X, \alpha )$ and $( Y, \beta )$ is defined to be the quotient space \\

(3.8)~~~ $ X \bigotimes_{\alpha, \beta} Y = Z / \approx_{\alpha, \beta} $ \\

with the mapping $\iota_{\alpha, \beta}$ induced through the inclusion (3.4) by the canonical quotient embedding
corresponding to (3.8), namely \\

(3.9)~~~ $ X \times Y \ni ( x, y ) \stackrel{\iota_{\alpha, \beta}}\longmapsto
                           x \bigotimes_{\alpha, \beta} y \in X \bigotimes_{\alpha, \beta} Y $ \\

where as in the usual case of tensor products, we denote by $x \bigotimes_{\alpha, \beta} y$, or simply $x \bigotimes y$,
the equivalence class of $( x, y )$. \\

Furthermore, the equivalence $\approx_{\alpha, \beta}$ is {\it compatible} with the semigroup structure $( Z, \gamma )$,
thus (3.8) has in fact the stronger form which gives a {\it commutative semigroup} structure on the resulting generalized
tensor product $X \bigotimes_{\alpha, \beta} Y$, namely \\

(3.10)~~~ $ ( X \bigotimes_{\alpha, \beta} Y, \gamma / \approx_{\alpha, \beta} ) =
                                                 ( Z, \gamma ) / \approx_{\alpha, \beta} $ \\

For simplicity, however, we shall write $\gamma$ instead of $\gamma / \approx_{\alpha, \beta}$. \\

In this way, the elements of $X \bigotimes_{\alpha, \beta} Y$ are all the expressions \\

(3.11)~~~ $ x_1 \bigotimes_{\alpha, \beta} y_1 ~\gamma~ x_2 \bigotimes_{\alpha, \beta} y_2 ~\gamma~
                                                   \ldots ~\gamma~ x_n \bigotimes_{\alpha, \beta} y_n $ \\

with $n \geq 1$ and $x_i \in X,~ y_i \in Y$, for $1 \leq i \leq n$. \\

Before going further, let us see when is the mapping (3.9) {\it injective}. A {\it necessary} condition is given by \\

{\bf Proposition 3.1.} \\

If the mapping $\iota_{\alpha, \beta}$ in (3.9), namely \\

(3.12)~~~ $ X \times Y \ni ( x, y ) \stackrel{\iota_{\alpha, \beta}}\longmapsto x
                                    \bigotimes_{\alpha, \beta} y \in X \bigotimes_{\alpha, \beta} Y $ \\

is {\it injective}, then the binary operations $\alpha$ and $\beta$ are {\it associative}. \\

{\bf Proof.} \\

We first show that \\

(3.13)~~~ $ \alpha $~ not associative $~ \Longrightarrow~ \iota_{\alpha, \beta} $~ not injective \\

Indeed, let $a, b, c \in X$, such that $d = \alpha ( \alpha ( a, b ), c ) \neq \alpha ( a, \alpha ( b, c ) ) = e$. Further,
let $y \in X$. Then in view of (3.6), we have \\

(3.14)~~~$ \begin{array}{l}
                 ( d, y ) = ( \alpha ( \alpha ( a, b ) ), c ), y )
                 \approx_{\alpha, \beta} ~( \alpha ( a, b ), y ) ~\gamma~ ( c, y ) \approx_{\alpha, \beta} \\ \\

                 ~~~~~~ \approx_{\alpha, \beta} ( a, y ) ~\gamma~ ( b, y ) ~\gamma~ ( c, y) \\ \\
                 ~~~~~~ \approx_{\alpha, \beta} ( a, y ) ~\gamma~ ( \alpha ( b, c ), y ) \approx_{\alpha, \beta} \\ \\

                 ~~~~~~ \approx_{\alpha, \beta} ( \alpha ( a, \alpha ( b, c )), y ) = ( e, y )
           \end{array} $ \\

hence $( d, y ) \approx_{\alpha, \beta} ( e, y )$, while obviously $( d, y ) \neq ( e, y )$. \\

In a similar manner, we also have \\

(3.15)~~~ $ \beta $~ not associative $~ \Longrightarrow~ \iota_{\alpha, \beta} $~ not injective \\

\hfill $\Box$ \\

The converse of Proposition 3.1. does {\it not} hold, as illustrated in \\

{\bf Example 3.1.} \\

The above definition contains as a particular case the usual tensor products of groups. And for Abelian groups one has \\

$~~~~~~ \mathbb{Z} / ( m ) \bigotimes_\mathbb{Z} \mathbb{Z} / ( n ) = \mathbb{Z} / ( d ) $ \\

for $m, n \in \mathbb{N}$, and $d$ the greatest common divisor of $m$ and $n$. Thus in particular \\

$~~~~~~ \mathbb{Z} / ( 2 ) \bigotimes_\mathbb{Z} \mathbb{Z} / ( 3 ) = 0 $

\hfill $\Box$ \\

Clearly, the binary operation $\gamma$ on $Z$ will canonically lead by this quotient operation to a {\it commutative} and
{\it associative} binary operation on $X \bigotimes_{\alpha, \beta} Y$, which for convenience is denoted by the same
$\gamma$, although this time it depends on $\alpha$ and $\beta$. \\

The customary and highly particular situation is when $X$ and $Y$ are semigroups, groups, or even vector spaces over some
field $\mathbb{K}$. In this case $\alpha, \beta$ and $\gamma$ are as usual denoted by +, that is, the sign of addition. \\

It is easy to note that in the construction of tensor products above, it is {\it not} necessary for $( X, \alpha )$ and $( Y, \beta )$ to be semigroups,
let alone groups, or for that matter, vector spaces. Indeed, it is sufficient that $\alpha$ and $\beta$ are arbitrary
binary operations on $X$ and $Y$, respectively. \\

Also, as seen above, $\alpha$ and $\beta$ need {\it not} be commutative either. However, the tensor product $X
\bigotimes_{\alpha, \beta} Y$, with the respective binary operation $\gamma$, will nevertheless be commutative and
associative. \\

It is important to note, [9], that the tensor products defined above have a {\it universality} property which is a natural
generalization of the corresponding well known one, see (A6.3), for usual tensor products. \\

{\bf Definition 3.1.} \\

Given two binary operations $\alpha : X \times X \longrightarrow X$ and $\beta : Y \times Y \longrightarrow Y$. An element $w \in
X \bigotimes_{\alpha, \beta} Y$ is called {\it entangled}, if and only if it is {\it not} of the form \\

(3.16)~~~ $ w = x \bigotimes_{\alpha, \beta} y $ \\

for some $x \in X$ and $y \in Y$. \\

{\bf Note 3.1.} \\

1) Since it was noted that the usual tensor products are particular cases of the tensor products defined in this section,
it follows that the definition of entanglement given above does indeed generalize the usual concept of entanglement. \\

2) It is important to note that generalized tensor products (3.8) can have an interest even when the corresponding
mappings (3.9) are not injective. Indeed, if for instance in such cases one still has the {\it strict} inclusion \\

(3.17)~~~ $ \iota_{\alpha, \beta} ( X \times Y ) \subsetneqq X \bigotimes_{\alpha, \beta} Y $ \\

then there are still {\it entangled} elements in $X \bigotimes_{\alpha, \beta} Y$, namely, those in the nonvoid set \\

(3.18)~~~ $ X \bigotimes_{\alpha, \beta} Y ~\setminus~ \iota_{\alpha, \beta} ( X \times Y ) $ \\

3) As seen in [10], tensor products can be defined in far more general ways than above, or for that matter, in the next
section 4. And with such far more general definitions there are plenty of cases when the mappings corresponding to (3.9)
will be injective.

\hfill $\Box$ \\

In the construction of tensor products above, we used the following easy to prove \\

{\bf Lemma 3.1.} \\

Let on a nonvoid set $E$ be given a family $( \equiv_i )_{i \in I}$ of {\it symmetric} binary relations. Further, let us
define on $E$ the binary relation $\approx$ as follows. For $a, b \in E$, we have $a \approx b$, if and only if $a = b$,
or there exists a finite sequence \\

$~~~~~~ a = c_0 \equiv_{i_0} c_1 \equiv_{i_1} c_2 \equiv_{i_2} \ldots \equiv_{i_{n-2}} c_{n-1} \equiv_{i_{n-1}} c_n = y $ \\

where $c_1, \ldots , c_{n-1} \in E$. \\

Then $\approx$ is an {\it equivalence} relation on $E$. \\ \\

{\bf 4. A General Concept of Entanglement} \\

We shall further generalize the concepts of tensor products and entanglement presented above. Namely, we shall consider
sets with arbitrary generators, that is, not necessarily associated with binary operations. \\

Let us start with the respective second generalization of the definition of tensor products. Given two structures
$( X, \psi )$ and $( Y, \varphi )$, where $\psi : {\cal P} ( X ) \longrightarrow {\cal P} ( X ),~ \varphi :
{\cal P} ( Y ) \longrightarrow {\cal P} ( Y )$ are arbitrary generators on $X$ and $Y$, respectively. Let us again denote
by $Z$ the set of all finite sequences of pairs \\

(4.1)~~~ $ ( x_1, y_1 ), \dots , ( x_n, y_n ) $ \\

where $n \geq 1$, while $x_i \in X,~ y_i \in Y$, with $1 \leq i \leq n$. Once more, we define on $Z$ the binary operation $\gamma$  simply by the
concatenation of the sequences (4.1). It follows that $\gamma$ is associative, therefore, each sequence (4.1) can be written as \\

(4.2)~~~ $ ( x_1, y_1 ), \dots , ( x_n, y_n ) =
                 ( x_1, y_1 ) ~\gamma~ ( x_2, y_2 ) ~\gamma~ \ldots ~\gamma~ ( x_n, y_n ) $ \\

where for $n = 1$, the right hand term is understood to be simply $( x_1, y_1 )$. Obviously, if $X$ or $Y$ have at least two elements, then $\gamma$ is
not commutative. \\

Thus we have \\

(4.3)~~~ $ Z = \left \{ ( x_1, y_1 ) ~\gamma~ ( x_2, y_2 ) ~\gamma~ \ldots ~\gamma~ ( x_n, y_n ) ~~
                             \begin{array}{|l}
                              ~ n \geq 1 \\ \\
                              ~ x_i \in X,~ y_i \in Y,~ 1 \leq i \leq n
                             \end{array} \right \} $ \\ \\

which obviously gives \\

(4.4)~~~ $ X \times Y \subseteq Z $ \\

Now we define on $Z$ an equivalence relation $\approx_{\psi, \varphi}$ as follows. Two sequences in (4.1) are equivalent, if and only if they are
identical, or each can be obtained from the other by a finite number of applications of the following operations : \\

(4.5)~~~ permute pairs $( x_i, y_i )$ within the sequence \\ \\

(4.6)~~~ replace $( x_1, y_1 ) ~\gamma~ ( x\,'_1, y_1 ) ~\gamma~ ( x_2, y_2 ) ~\gamma~ \ldots ~\gamma~ ( x_n, y_n )$ \\
         \hspace*{1.3cm} with $( \alpha ( x_1, x\,'_1), y_1 ) ~\gamma~ ( x_2, y_2 ) ~\gamma~ \ldots , ( x_n, y_n )$,
         or vice-versa, \\
         \hspace*{1.3cm} where $\alpha$ is a binary operation on $X$ which is compatible
         \hspace*{1.3cm} with $\psi$ \\ \\

(4.7)~~~ replace $( x_1, y_1 ) ~\gamma~ ( x_1, y\,'_1 ) ~\gamma~ ( x_2, y_2 ) ~\gamma~ \ldots ~\gamma~ ( x_n, y_n )$ \\
         \hspace*{1.3cm} with $( x_1, \beta ( y_1, y\,'_1 ) ) ~\gamma~ ( x_2, y_2 ) ~\gamma~ \ldots ~\gamma~
         ( x_n, y_n )$, or vice-versa, \\
         \hspace*{1.3cm} where $\beta$ is a binary operation on $Y$ which is compatible \\
         \hspace*{1.3cm} with $\varphi$ \\ \\

In view of 2) in Remark 1.1. above, there always exists a binary operation $\alpha$ compatible with $\psi$, and also a binary operation $\beta$
compatible with $\varphi$. \\

We note again that, in view of Lemma 3.1., the binary relation $\approx_{\psi, \varphi}$ defined above is indeed an
equivalence relation on $Z$. \\

Finally, the {\it tensor product} of $( X, \psi )$ and $( Y, \varphi )$ is defined to be the quotient space \\

(4.8)~~~ $ X \bigotimes_{\psi, \varphi} Y = Z / \approx_{\psi, \varphi} $ \\

with the mapping $\iota_{\psi, \varphi}$ induced through the inclusion (4.4) by the canonical quotient embedding
corresponding to (4.8), namely \\

(4.9)~~~ $ X \times Y \ni ( x, y ) \stackrel{\iota_{\psi, \varphi}}\longmapsto
                         x \bigotimes_{\psi, \varphi} y \in X \bigotimes_{\psi, \varphi} Y $ \\

where as in the usual case of tensor products, we denote by $x \bigotimes_{\psi, \varphi} y$, or simply  $x \bigotimes y$, the equivalence class of
$( x, y ) \in X \times Y \subseteq Z$. \\

Again, similar with Proposition 3.1., one can obtain conditions for the mapping $\iota_{\psi, \varphi}$ in (4.9) to be
{\it injective}. \\

Obviously, the binary operation $\gamma$ on $Z$ will canonically lead by this quotient operation to a {\it commutative}
and {\it associative} binary operation on $X \bigotimes_{\psi, \varphi} Y$, which for convenience is denoted by the same
$\gamma$, although in view of (4.8), this time it depends on $\psi$ and $\varphi$. In this way, the elements of
$X \bigotimes_{\alpha, \beta} Y$ are all the expressions \\

(4.10)~~~ $ x_1 \bigotimes y_1 ~\gamma~ x_2 \bigotimes y_2 ~\gamma~ \ldots ~\gamma~ x_n \bigotimes y_n $ \\

with $n \geq 1$ and $x_i \in X,~ y_i \in Y$, for $1 \leq i \leq n$. \\

It is important to note, [9], that the tensor products defined above have a {\it universality} property which is a natural,
albeit rather surprising generalization of the corresponding well known one, see (A6.3), for usual tensor products. \\

{\bf Definition 4.1.} \\

An element $w \in X \bigotimes_{\psi, \varphi} Y$ is called {\it entangled}, if and only if it is {\it not} of the form \\

(4.11)~~~ $ w = x \bigotimes_{\psi, \varphi} y $ \\

for some $x \in X$ and $y \in Y$. \\

We conclude by \\

{\bf Theorem 4.1.} \\

The tensor products constructed in section 3 above are particular cases of those in this section. \\

{\bf Proof.} \\

Let be given two structures $( X, \alpha )$ and $( Y, \beta )$, where $\alpha : X \times X \longrightarrow X,~ \beta : Y \times Y \longrightarrow Y$ are
arbitrary binary operations on $X$ and $Y$, respectively. Then as in (1.5), we associate with them the generators $\psi_\alpha$ and $\psi_\beta$ on $X$
and $Y$, respectively. \\

We show now that, for $z, z\,' \in Z$, we have the implication \\

(4.12)~~~ $ z \approx_{\alpha, \beta} z\,' ~~~\Longrightarrow~~~ z \approx_{\psi_\alpha, \psi_\beta} z\,' $ \\

Indeed, it is sufficient to prove that (3.6) implies (4.6), and (3.7) implies (4.7). And clearly, in both implications we can assume $n = 1$ without
loss of generality. \\

Let us therefore be given $x, x\,' \in X,~ y \in Y$. If we assume (3.6), then we obtain \\

(4.13)~~~ $ ( \alpha ( x, x\,' ), y ) \approx_{\alpha, \beta} ( x, y ) ~\gamma~ ( x\,' y ) $ \\

But in view of 1) in Remark 1.1., $\alpha$ is compatible with $\phi_\alpha$, thus (4.13) gives \\

(4.14)~~~ $ ( \alpha ( x, x\,' ), y ) \approx_{\psi_\alpha, \psi_\beta} ( x, y ) ~\gamma~ ( x\,' y ) $ \\

and the implication (3.6) $\Longrightarrow$ (4.6) is proved. The proof of the implication (3.7) $\Longrightarrow$ (4.7) is
similar. \\

{\bf Note 4.1.} \\

1) In view of Theorem 4.1., it follows that the concept of entanglement in Definition 4.1. contains as a particular case that in Definition 3.1., and
therefore, also the usual concept of entanglement. \\

2) The interest in the general concept of entanglement in Definition 4.1. is, among others, in the fact that it is {\it no longer} confined within an
algebraic context. In this way, this paper shows that entanglement can, so to say, be {\it de-entangled} not only from tensor products of vector spaces,
groups or semigroups, but also more generally, altogether from algebra. \\

3) The usual concept of entanglement in tensor products is in fact given by a {\it negation}, that is, the inexistence of a certain kind of specific
representation. Consequently, any extension and deepening of that concept is likely to open up a large variety of meaningful possibilities for additional
forms of entanglement. Indeed, such extensions and deepening, based as above, on generalizations of the usual concept of tensor product, are less likely
to enlarge the specific conditions defining non-entanglement, than are likely to enlarge the realms of negations of those specific conditions, thus
enlarging the possibilities for entanglement. \\

4) Quantum physics arguments of quite some concern related to the usual tensor product based concept of entanglement were recently presented in [7]. And
they indicate what may be seen as a lack of {\it ontological robustness} of that concept. As an effect, one may expect that what appears to be
entanglement in terms of usual tensor products may in fact correspond to considerably deeper aspects. In this regard, the old saying that "the whole is
{\it more} than the sum of its parts" may in fact mean that what is involved in that "more" can correspond to very different things, depending on the
situation. \\

5) Applications of the general concept of entanglement in Definition 4.1. are to be presented
in a subsequent paper. \\ \\

{\bf Appendix} \\

For convenience, we recall here certain main features of the usual tensor product of vector
spaces, and relate them to certain properties of Cartesian products. \\

Let $\mathbb{K}$ be a field and $E,F, G$ vector spaces over $\mathbb{K}$. \\

{\bf A1. Cartesian Product of Vector Spaces} \\

Then $E \times F$ is the vector space over $\mathbb{K}$ where the operations are given by \\

$~~~~~~ \lambda ( x, y ) + \mu ( u, v ) ~=~ ( \lambda x + \mu u, \lambda y + \mu v ) $ \\

for any $x, y \in E,~ u, v \in F,~ \lambda, \mu \in \mathbb{K}$. \\ \\

{\bf A2. Linear Mappings} \\

Let ${\cal L} ( E, F )$ be the set of all mappings \\

$~~~~~~ f : E ~\longrightarrow~ F $ \\

such that \\

$~~~~~~ f ( \lambda x + \mu u ) ~=~ \lambda f ( x ) + \mu f ( u ) $ \\

for $u, v \in E,~ \lambda, \mu \in \mathbb{K}$. \\ \\

{\bf A3. Bilinear Mappings} \\

Let ${\cal L} ( E, F; G )$ be the set of all mappings \\

$~~~~~~ g : E \times F ~\longrightarrow~ G $ \\

such that for $x \in E$ fixed, the mapping $F \ni y \longmapsto g ( x, y ) \in G$ is linear in
$y$, and similarly, for $y \in F$ fixed, the mapping $E \ni x \longmapsto g ( x, y ) \in G$ is
linear in $x \in E$. \\

It is easy to see that \\

$~~~~~~ {\cal L} ( E, F; G ) ~=~ {\cal L} ( E, {\cal L} ( F, G ) ) $ \\ \\

{\bf A4. Tensor Products} \\

The aim of the tensor product $E \bigotimes F$  is to establish a close connection between the
{\it bilinear} mappings in ${\cal L} ( E, F; G )$ and the {\it linear} mappings in
${\cal L} ( E \bigotimes F , G )$. \\

Namely, the {\it tensor product} $E \bigotimes F$ is : \\

(A4.1)~~~ a vector space over $\mathbb{K}$, together with \\

(A4.2)~~~ a bilinear mapping $t : E \times F ~\longrightarrow~ E \bigotimes F$, such that we \\
          \hspace*{1.6cm} have the following : \\ \\

{\bf UNIVERSALITY PROPERTY} \\

$\begin{array}{l}
    ~~~~~~~~~~ \forall~~~ V ~\mbox{vector space over}~ \mathbb{K},~~
                   g \in {\cal L} ( E, F; V ) ~\mbox{bilinear mapping}~ : \\ \\
    ~~~~~~~~~~ \exists~ !~~ h \in {\cal L} ( E \bigotimes F, V )
                                         ~\mbox{linear mapping}~ : \\ \\
    ~~~~~~~~~~~~~~~~ h \circ t ~=~ g
  \end{array} $ \\ \\

or in other words : \\

(A4.3)~~~ the diagram commutes

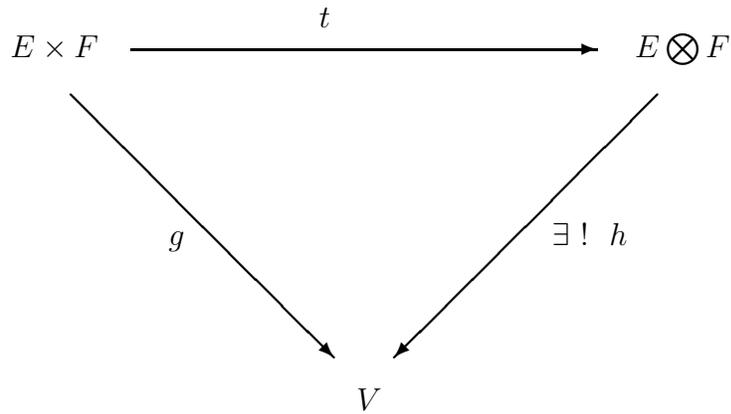
\begin{math}
\setlength{\unitlength}{1cm}
\thicklines
\begin{picture}(13,7)

\put(0.9,5){$E \times F$}
\put(2.5,5.1){\vector(1,0){6.2}}
\put(9.2,5){$E \bigotimes F$}
\put(5,5.4){$t$}
\put(1.7,4.5){\vector(1,-1){3.5}}
\put(9.5,4.5){\vector(-1,-1){3.5}}
\put(5.5,0.3){$V$}
\put(3,2.5){$g$}
\put(8.1,2.5){$\exists~!~~ h$}

\end{picture}
\end{math}

and \\

(A4.4)~~~ the tensor product $E \bigotimes F$ is {\it unique} up to vector \\
          \hspace*{1.6cm} space isomorphism. \\

Therefore we have the {\it injective} mapping \\

$~~~~~~ {\cal L} ( E, F; V ) \ni g ~\longmapsto~ h \in {\cal L} ( E \bigotimes F, V )
                                                    ~~~~\mbox{with}~~~~ h \circ t ~=~ g $ \\

The converse mapping \\

$~~~~~~ {\cal L} ( E \bigotimes F, V ) \ni h ~\longmapsto~
                         g ~=~ h \circ t \in {\cal L} ( E, F; V ) $ \\

obviously exists. Thus we have the {\it bijective} mapping \\

$~~~~~~ {\cal L} ( E \bigotimes F, V ) \ni h ~\longmapsto~
                         g ~=~ h \circ t \in {\cal L} ( E, F; V ) $ \\ \\

{\bf A5. Lack of Interest in ${\cal L} ( E \times F, G )$} \\

Let $f \in {\cal L} ( E \times F, G )$ and $( x, y ) \in E \times F$, then $( x, y ) = ( x, 0 )
+ ( 0, y )$, hence \\

$~~~~~~ f ( x, y ) ~=~ f ( ( x, 0 ) + ( 0, y ) ) ~=~ f ( x, 0 ) ~+~ f ( 0, y ) $ \\

thus $f ( x, y )$ depends on $x$ and $y$ in a {\it particular} manner, that is, separately on
$x$, and separately on $y$. \\ \\

{\bf A6. Universality Property of Cartesian Products} \\

Let $X, Y$ be two nonvoid sets. Their cartesian product is : \\

(A6.1)~~~ a set $X \times Y$, together with \\

(A6.2)~~~ two projection mappings~ $p_X : X \times X ~\longrightarrow~ X, \\
          \hspace*{1.6cm} p_Y : X \times Y ~\longrightarrow~ Y$, such that we have the
          following : \\ \\

{\bf UNIVERSALITY PROPERTY} \\

$\begin{array}{l}
    ~~~~~~~~~~ \forall~~~ Z ~\mbox{nonvoid set},~~ f : Z ~\longrightarrow~ X,~~
                          g : Z ~\longrightarrow~ Y~ : \\ \\
    ~~~~~~~~~~ \exists~ !~~ h  : Z ~\longrightarrow~ X \times Y~ : \\ \\
    ~~~~~~~~~~~~~~~~ f ~=~ p_X \circ h,~~~ g ~=~ p_Y \circ h
  \end{array} $ \\ \\

or in other words : \\

(A6.3)~~~ the diagram commutes

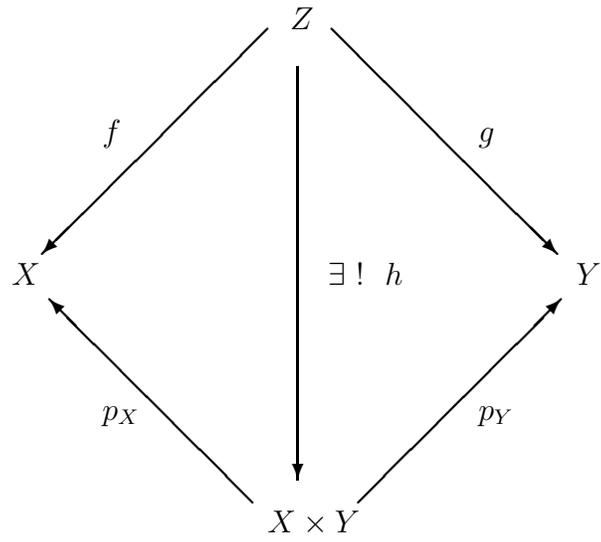
\begin{math}
\setlength{\unitlength}{1cm}
\thicklines
\begin{picture}(20,9)

\put(6.5,3.6){$\exists~ ! ~~ h$}
\put(6.1,6.5){\vector(0,-1){5.5}}
\put(6,7){$Z$}
\put(3.5,5.5){$f$}
\put(5.7,7){\vector(-1,-1){3}}
\put(8.5,5.5){$g$}
\put(6.55,7){\vector(1,-1){3}}
\put(2.3,3.6){$X$}
\put(9.8,3.6){$Y$}
\put(3.5,1.8){$p_X$}
\put(5.5,0.7){\vector(-1,1){2.7}}
\put(8.5,1.8){$p_Y$}
\put(6.9,0.7){\vector(1,1){2.7}}
\put(5.7,0.3){$X \times Y$}

\end{picture}
\end{math} \\ \\

{\bf A7. Cartesian and Tensor Products seen together} \\

\newpage

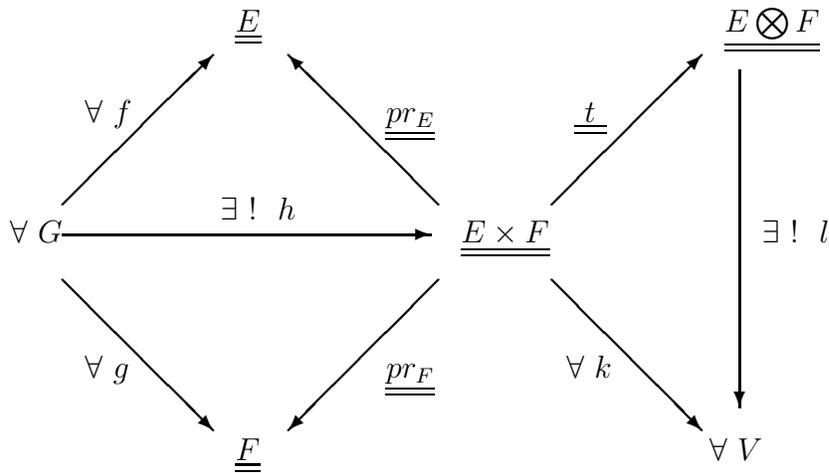
\begin{math}
\setlength{\unitlength}{1cm}
\thicklines
\begin{picture}(30,10)

\put(-0.2,6){$\forall~G$}
\put(0.5,6.5){\vector(1,1){2}}
\put(0.8,7.6){$\forall~f$}
\put(2.8,8.8){$\underline{\underline{E}}$}
\put(0.5,5.5){\vector(1,-1){2}}
\put(0.8,4.2){$\forall~g$}
\put(2.8,3.1){$\underline{\underline{F}}$}
\put(5.8,6){$\underline{\underline{E \times F}}$}
\put(5.5,6.5){\vector(-1,1){2}}
\put(4.8,7.6){$\underline{\underline{pr_E}}$}
\put(5.5,5.5){\vector(-1,-1){2}}
\put(4.8,4.2){$\underline{\underline{pr_F}}$}
\put(0.5,6.1){\vector(1,0){4.9}}
\put(2.6,6.3){$\exists~ !~~ h$}
\put(7,6.5){\vector(1,1){2}}
\put(7.3,7.6){$\underline{\underline{~t~}}$}
\put(7,5.5){\vector(1,-1){2}}
\put(7.2,4.2){$\forall~k$}
\put(9.3,8.8){$\underline{\underline{E \bigotimes F}}$}
\put(9.1,3.1){$\forall~V$}
\put(9.5,8.3){\vector(0,-1){4.5}}
\put(9.8,6){$\exists~ ! ~~ l$}

\end{picture}
\end{math}

{\bf Acknowledgment} \\

Grateful thanks to Michiel Hazewinkel for Example 3.1., and to my young collaborators Claudia Zander and Gusti van Zyl, for inspiring discussions and
the correction of some errors. \\ \\

\end{document}